\documentclass[twocolumn,nofootinbib,aps,showpacs,prx,preprintnumbers]{revtex4-2}

\usepackage{graphicx}
\usepackage{dcolumn}
\usepackage{epsfig}
\usepackage{epstopdf}
\usepackage{amssymb}
\usepackage{amsfonts}
\usepackage{xcolor}
\usepackage{float}
\usepackage{amsmath}
\usepackage{hyperref}
\usepackage{physics}
\usepackage{comment}
\usepackage[caption=false]{subfig}
\usepackage{soul}
\usepackage[bottom]{footmisc}

\usepackage{mathtools}
\DeclareMathAlphabet\mathbfcal{OMS}{cmsy}{b}{n}

\hypersetup{
	colorlinks = true,
	linkcolor = blue,
	citecolor = blue,
	urlcolor = blue
}
\allowdisplaybreaks
\setstcolor{red}

\begin{document}

\title{
Doppler-Enhanced Quantum Magnetometry with thermal Rydberg atoms
}

\author{Shovan Kanti Barik}
\author{Silpa B S}
\author{M Venkat Ramana}
\author{Shovan Dutta}
\author{Sanjukta Roy}
\email{sanjukta@rri.res.in}

\affiliation{Raman Research Institute, C. V. Raman Avenue, Sadashivanagar, Bangalore 560080, India}

\begin{abstract}
We report experimental measurements showing how one can combine quantum interference and thermal Doppler shifts at room temperature to detect weak magnetic fields. We pump ${}^{87}$Rb atoms to a highly-excited, Rydberg level using a probe and a coupling laser, leading to narrow transmission peaks of the probe due to destructive interference of transition amplitudes, known as Electromagnetically Induced Transparency (EIT). While it is customary in such setups to use counterpropagating lasers to minimize the effect of Doppler shifts, here we show, on the contrary, that one can harness Doppler shifts in a copropagating arrangement to produce an enhanced response to a magnetic field. In particular, we demonstrate an order-of-magnitude bigger splitting in the transmission spectrum as compared to the counterpropagating case. We explain and generalize our findings with theoretical modeling and simulations based on a Lindblad master equation. Our results pave the way to using quantum effects for magnetometry in readily deployable room-temperature platforms.
\end{abstract}
\maketitle

\section{Introduction}
In recent years Rydberg atoms have opened up novel avenues and provided a powerful and versatile platform for exploring both fundamental problems of many-body physics through quantum simulation \cite{2021_Browaeys} and important technological applications \cite{Adams_2019} such as quantum information processing \cite{Saffman_2010}, photon-photon quantum gates \cite{2019_Rempe}, precision measurements using quantum sensing \cite{2021_Debatin, 2021_Kunz}, and quantum communication using cold Rydberg atoms as single-photon sources \cite{2018_Pfau, 2013_Pfau}.

Quantum sensors \cite{2017_RMP} utilize quantum properties such as interference, entanglement, or squeezed states to precisely measure a physical quantity. In particular, precision magnetometry has a wide variety of applications in archaeology, space explorations, geophysics, detection of brain activity, and detection of mineralisations. While cryogenically cooled superconducting devices set the bar for sensing ultraweak magnetic fields, room-temperature vapor-cell experiments are much more convenient to deploy for practical applications \cite{2012_Pfau} due to the simplified experimental system and non-requirement of atom cooling or ultra-high vacuum. Here we use the quantum interference phenomenon of Electromagnetically Induced Transparency (EIT) \cite{EIT_Review_Marangos}, in conjunction with thermal Doppler shifts, to demonstrate enhanced response to magnetic fields of a room-temperature atomic vapor. 

In a typical Rydberg EIT configuration a probe beam couples the ground state to an intermediate, excited state of finite lifetime, which in turn is coupled to a highly-excited, metastable Rydberg state by a coupling beam. At two-photon resonance the transition amplitudes from the ground and Rydberg levels to the intermediate level interfere destructively, leading to vanishing absorption of the probe beam over a narrow transparency window. EIT has important applications in slowing of light \cite{1999_Hau}, quantum memory \cite{Ma_2017}, light storage \cite{2001_Lukin}, and accurate, nondestructive mapping of Rydberg levels \cite{2007_Mohapatra, 2022_Rydberg_TF}. 

At room temperature the motion of atoms leads to Doppler shifts for both the probe and the coupling lasers. When the two are counterpropgating and have the same frequency, these Doppler shifts cancel out in the two-photon resonance, and one obtains a narrow EIT signal. Generically, the frequencies are different and the cancellation is only partial \cite{Urvoy_2013}. However, it is still conventional to use the counterpropagating arrangement to minimize Doppler broadening. Magnetometry using this EIT configuration have been explored in Refs.~\cite{2017_Cheng, 2018_Zhang, 2022_Opt_Ryd_EIT, 2016_pol_spectra, 2017_Spreeuw}. 

In contrast, we demonstrate an order of magnitude enhancement of the magnetic-field response in a copropagating arrangement for various polarizations of the two lasers. Furthermore, we show that such an enhancement is obtained whenever the Zeeman shift of the intermediate level is larger (or smaller) than those of both the ground and the Rydberg levels, and the enhancement factor is controlled by the ratio $\omega_p / \omega_c$ where $\omega_{p(c)}$ is the probe (coupling) frequency. These findings have important applications toward quantum sensing of magnetic fields using thermal Rydberg atoms.

\section{Experimental setup and methods}
\label{sec:Expt_setup}
\begin{figure}
    \centering
    \includegraphics[width=\columnwidth]{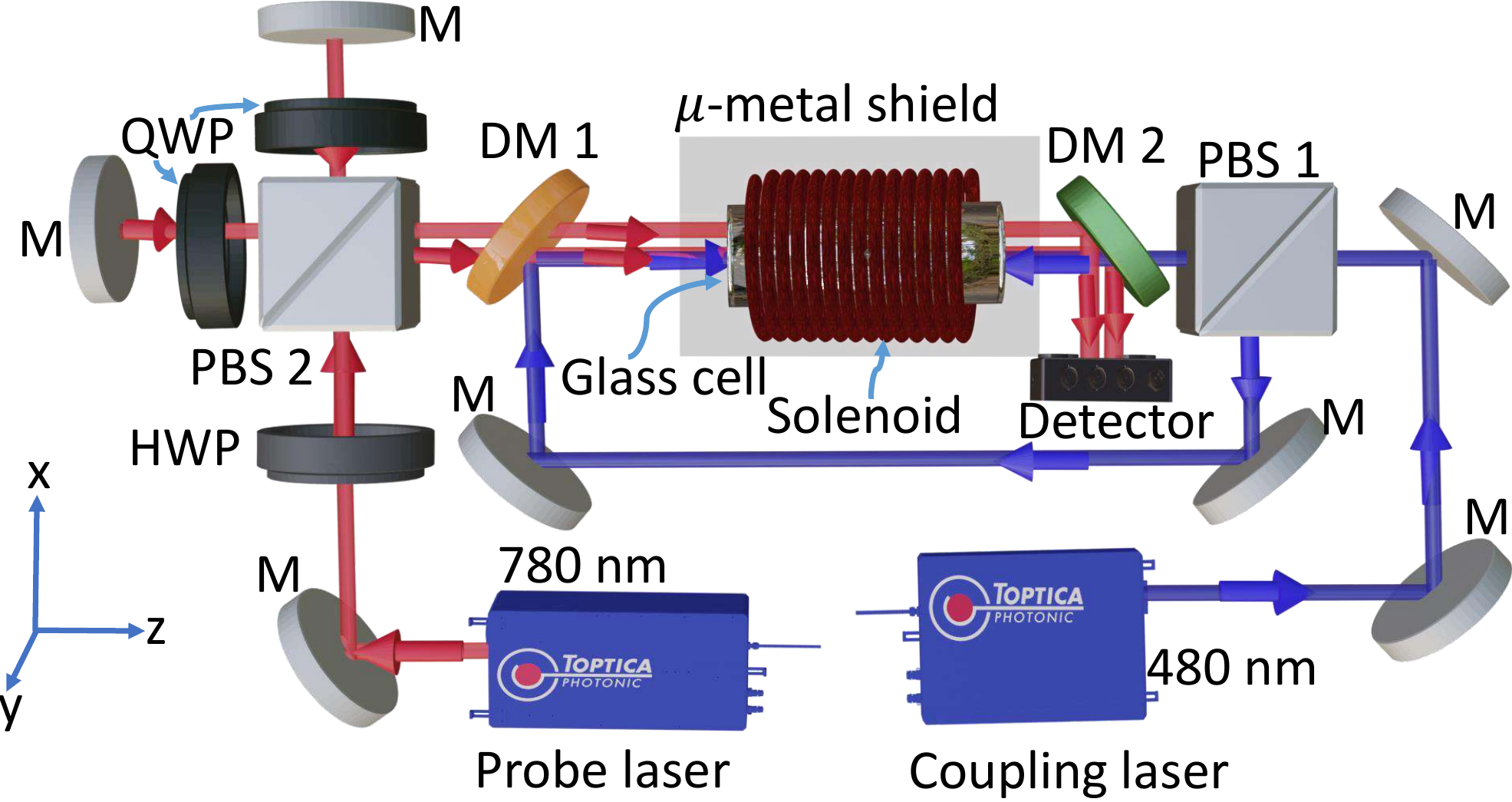}
    \caption{
    Schematic diagram of the experimental setup. Red beam represents the probe laser (780 nm) which is split into two parts by using a polarizing beamsplitter (PBS 2) and focused onto a homemade balanced detector after traveling through the vapor cell. Blue beam is divided into two coupling beams (480 nm) using a polarizing beamsplitter (PBS 1), which are then counterpropagated and overlapped with the probe beam using two dichoric mirrors (DMs). The vapor cell is placed along the axis of a solenoid coil and magnetically shielded by $\mu$-metal. M: mirror; PBS: polarizing beam splitter; HWP: half wave plate; QWP: quarter wave plate.
    } 
    \label{fig:setup}
\end{figure}

\begin{figure}
    \centering
    \includegraphics[width=\columnwidth]{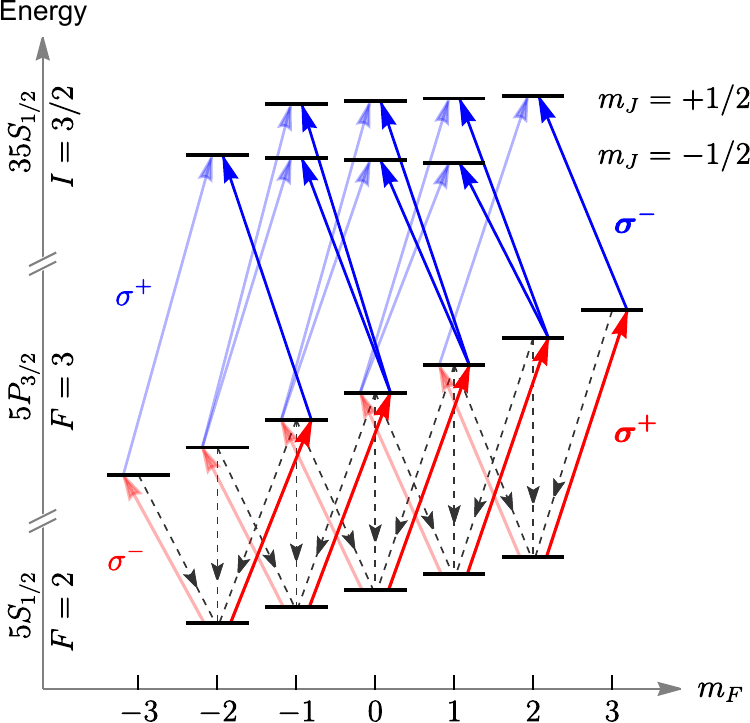}
    \caption{
    Energy-level diagram showing the ground manifold $\{5S_{1/2}, F=2\}$, the excited manifold $\{5P_{3/2}, F'=3\}$, and the Rydberg manifold $\{35S_{1/2}, I=3/2\}$ of ${}^{87}$Rb in the presence of a magnetic field. The red and blue arrows denote coherent pumping by the probe and coupling lasers, respectively, for $\sigma^{\pm}$ polarizations. The dashed lines represent incoherent decay of the excited states back to the ground manifold.
    }
    \label{fig:energylevels}
\end{figure}

The schematic diagram of the experimental setup is shown in Fig.~\ref{fig:setup}. For the Rydberg EIT we utilize three hyperfine manifolds $\{5S_{1/2}, F=2\} \rightarrow \{5P_{3/2}, F'=3\} \rightarrow \{35S_{1/2}, F''=2\}$ of $^{87}$Rb, as shown in  Fig.~\ref{fig:energylevels}. The hyperfine states of the Rydberg level $\{35S_{1/2},F''=2 \}$ are mixed with the $F''=1$ hyperfine sector at magnetic field $B \gtrsim 1$ G, resulting in 8 sublevels. The probe beam is  derived from an External Cavity Diode Laser (ECDL) (Toptica DL 100) and tuned to the D2 atomic transition $\{5S_{1/2}, F=2\} \rightarrow \{5P_{3/2}, F' =3\}$ at a wavelength of 780 nm. The coupling beam is derived from a tunable frequency-doubled laser (Toptica TA SHG pro) and tuned to the transition $\{5P_{3/2}, F'=3\} \rightarrow \{35S_{1/2}, F''=2\}$ at a wavelength of 480 nm by tuning the seed laser of the frequency-doubled laser to 960 nm. The frequencies of the seed laser and the frequency-doubled coupling laser are monitored using a commercial wavelength meter based on the Fizeau interferometer (HighFinesse WS8-2), having an absolute accuracy of 2 MHz and a frequency resolution of 200 kHz. In the experiment the probe laser is scanned across the $\{5S_{1/2}, F=2\} \rightarrow \{5P_{3/2}, F'=3\}$ transition and the coupling laser is kept fixed at the $\{5P_{3/2}, F'=3\} \rightarrow \{35S_{1/2}, F''=2\}$ transition.

As shown in Fig.~\ref{fig:setup}, the laser beam at 780 nm is split into two parts with a polarizing beam splitter (PBS 2), one used as a probe and the other as a reference beam. Both the probe and the reference beams with a Gaussian radius of  0.7 mm are  collimated and aligned through a Rb vapor cell of length 75 mm. The coupling laser beam is divided into two parts using a polarizing beamsplitter (PBS 1), with one aligned counterpropagating with the probe and the other aligned copropagating with the probe. The Gaussian radius of these two beams are  1.2 mm and 0.7 mm, respectively. The reference beam is propagated through the vapor cell without overlapping with any of the coupling beams to provide a Doppler-broadened background spectrum. The absorption spectra of the probe and the reference beams are recorded on a homemade  balanced photodiode, and the difference signal is  obtained to  record the Rydberg EIT signal on a flat background with a good signal-to-noise ratio.

A solenoid  coil is used to generate magnetic field along its axis (along z-axis in  Fig.~\ref{fig:setup}). The magnetic field along the axis of the solenoid coil is measured with a Gaussmeter Model 460 3-channel from Lake Shore Cryptonics. The magnetic field gradient of the solenoid coil with current is measured as 33.616 Gauss/Amp. The Rb vapor cell is placed along the axis of the solenoid coil where the magnetic field is almost uniform across the length of the vapor cell. We vary the magnetic field from 0 to 10 Gauss. The measured standard deviation of magnetic field along the vapor cell is around 0.2 Gauss at 10 Gauss magnetic field. The vapor cell and the solenoid are enclosed in a $\mu$-metal magnetic shield to prevent influence of ambient stray magnetic fields on the Rydberg EIT signals. Three layers of $\mu$-metal sheets are used to reduce the ambient stray magnetic field below 0.05 Gauss.

\section{\label{sec:theory}Theoretical Model}
The absorption spectra result from the complex susceptibility at the probe frequency in the steady state arising from the optical drive and the spontaneous decay of the excited states, as sketched in Fig.~\ref{fig:energylevels}. We obtain the steady state from a Lindblad master equation governing the density matrix of an atom, and average the resulting susceptibility over a thermal distribution.  

The ground manifold consists of 5 Zeeman sublevels $|m_F^g\rangle$ with energies $\varepsilon(m_F^g) = g^g_F m^g_F \mu_B B$, where $B$ is the magnetic field, $\mu_B$ is the Bohr magneton, $g_F^g$ is the Land{\'e}-$g$ factor, and $m_F^g \in \{-2,\dots,2\}$. Similarly, the excited manifold is split into 7 sublevels $|m_F^e\rangle$ with energies $\varepsilon(m_F^e) = \varepsilon_{g,e} + g^e_F m^e_F \mu_B B$, where $m_F^e \in \{-3,\dots,3\}$ and $\varepsilon_{g,e}$ is the energy from the ground level at $B=0$. In the Rydberg manifold the splittings for $B \sim 1$G are large enough to mix the hyperfine levels $F''=1$ and $F''=2$, decoupling the nuclear spin $I=3/2$ from the electronic angular momentum $J=1/2$. This Paschen-Back effect gives 8 sublevels $|m_I^r, m_J^r \rangle$ with energies $\varepsilon(m_I^r, m_J^r) = \varepsilon_{g,r} + A_{\text{hf}} m_I^r m_J^r + (g_I^r m_I^r + g_J^r m_J^r) \mu_B B$, where $\varepsilon_{g,r}$ is the reference from the ground level, $A_{\text{hf}}$ is the hyperfine interaction energy, $m_I^r \in \{\pm 1/2, \pm 3/2\}$, and $m_J^r \in \{\pm 1/2\}$. The Land{\'e}-$g$ factors are related to the angular momentum quantum numbers, which give $g_F^g = 1/2$, $g_F^e = 2/3$, and $g_J^r = 2$. We also have $A_{\text{hf}} = 2\pi\hbar \times 573$ kHz \cite{2013_Spreeuw} and $g_I^r = -0.001$ \cite{2018_OE} from prior experimental measurements. Hence, the largest contribution to the Rydberg splitting comes from $g^r_J$. Note the splittings are in MHz, whereas the gaps $\varepsilon_{g,e}$ and $\varepsilon_{g,r}$ are in PHz.

The coupling between the energy levels is given by the Hamiltonian $- \hat{\mathbf{d}}.\mathbf{E}(t)$, where $\hat{\mathbf{d}} \coloneqq -e \hat{\mathbf{r}}$ is the dipole operator and $\mathbf{E}(t)$ is the net electric field of the lasers. The latter can be expressed as $\mathbf{E}(t) = \text{Re}[\mathbfcal{E}_p e^{-{\rm i} \omega_p t} + \mathbfcal{E}_c e^{-{\rm i} \omega_c t}]$, where $\mathbfcal{E}_{p(c)}$ is the complex amplitude and $\omega_{p(c)} \sim$ PHz is the frequency of the probe (coupling) laser. To extract the time-averaged dynamics we move to a rotating frame with the transformation $\smash{\hat{U} = e^{{\rm i} \omega_p t \hat{P}_e + {\rm i} (\omega_p + \omega_c) t \hat{P}_r}}$, where $\smash{\hat{P}_{e(r)}}$ projects onto the excited (Rydberg) manifold. The transformed Hamiltonian $\hat{H}_{\circ}$ is given by $\smash{\hat{U} ({\rm i} \partial_t - \hat{H}) \hat{U}^{\dagger}} = {\rm i} \partial_t - \hat{H}_{\circ}$. Averaging out the oscillating terms in $\hat{H}_{\circ}$ yields a time-independent Hamiltonian with the rescaled energies $\varepsilon_{\circ}(m_F^g) = \varepsilon (m_F^g)$, $\varepsilon_{\circ}(m_F^e) =  g^e_F m^e_F \mu_B B - \Delta_p$, and $\varepsilon_{\circ}(m_I^r, m_J^r) = (g_I^r m_I^r + g_J^r m_J^r) \mu_B B + A_{\text{hf}} (m_I^r m_J^r - 3/4) - \Delta_p - \Delta_c$, where $\Delta_p \coloneqq \hbar \omega_p - \varepsilon_{g,e}$ is the detuning of the probe laser and $\Delta_c \coloneqq \hbar \omega_c + \varepsilon_{g,e} - \varepsilon_{g,r} - (3/4) A_{\text{hf}}$ is the detuning of the coupling laser, defined with respect to the $F''=2$ Rydberg level at $B=0$. The Rabi frequencies are given by $\Omega(m_F^g, m_F^e) = - \langle m_F^e | \smash{\hat{\mathbf{d}}.\mathbfcal{E}_p} | m_F^g \rangle$ and $\Omega(m_F^e, m_I^r, m_J^r) = - \langle \{m_I^r, m_J^r\} | \smash{\hat{\mathbf{d}}.\mathbfcal{E}_c} | m_F^e \rangle$, which lead to selection rules depending on the polarization of $\mathbfcal{E}_p$ and $\mathbfcal{E}_c$. In particular, for circular polarization $\sigma^{\pm}$ perpendicular to $\mathbf{B}$, $m_F$ changes only by $\pm 1$ (Fig.~\ref{fig:energylevels}).

The optical drive competes with the spontaneous decay of the excited states back to the ground manifold, which occurs over a lifetime of $\sim 0.1\; \mu$s. Radiative decay can happen through all polarization channels, so $m_F$ changes by $0$ or $\pm 1$ with different branching ratios (Fig.~\ref{fig:energylevels}). These individual decay rates as well as the dipole matrix elements entering the Rabi frequencies are independent of the drive parameters and are calculated using the \texttt{ARC} library \cite{vsibalic2017arc}. Note that the Rydberg levels have a long lifetime ($\sim 20 \; \mu$s \cite{branden2009radiative}) which does not significantly alter the steady state.

Under the usual Born-Markov approximation \cite{daley2014quantum} the driven-dissipative dynamics are governed by a Lindblad master equation
\begin{equation*}
    {\rm d} \hat{\rho}_{\circ} / {\rm d}t = 
    {\rm i} [\hat{\rho}_{\circ}, \hat{H}_{\circ}/\hbar] 
    + \sum\nolimits_k \gamma_k \big(
    \hat{L}_k \hat{\rho}_{\circ} \hat{L}_k^{\dagger} 
    - \{\hat{L}_k^{\dagger} \hat{L}_k, \hat{\rho}_{\circ}\}/2 
    \big) ,
\end{equation*}
where $\hat{\rho}_{\circ}$ is the density matrix in the rotating frame and $\hat{L}_k$ describe the transitions $|m_F^g \rangle \langle m_F^e|$ with corresponding decay rates $\gamma_k$. We find a unique steady state for $\hat{\rho}_{\circ}$ using exact diagonalization. Moving back to the lab frame, $\hat{\rho} = \hat{U}^{\dagger}(t) \hat{\rho}_{\circ} \hat{U}(t)$, one finds that the coherences between ground and excited (excited and Rydberg) manifold oscillate at frequency $\omega_p$ ($\omega_c$).

The probe absorption originates from a complex refractive index $n(\omega_p) = \sqrt{1 + \chi(\omega_p)}$ \cite{steck2007quantum}, where the susceptibility $\chi$ is defined in terms of the dipole moment per unit volume, $N \langle \smash{\hat{\mathbf{d}}} \rangle = \varepsilon_0 \chi \mathbf{E}$. Here $N$ is the atom number density and $\varepsilon_0$ is the permittivity of free space. Writing $\langle \smash{\hat{\mathbf{d}}} \rangle = \text{Tr}[ \smash{\hat{\mathbf{d}}} \hat{\rho}(t)]$ and comparing Fourier coefficients yield $\chi(\omega_p) = [2N/(\varepsilon_0 \mathcal{E}_p)]\; \text{Tr}\big(\smash{\hat{P}_g \hat{d}_p \hat{P}_e \hat{\rho}_{\circ}} \big)$, where $\smash{\hat{d}_p}$ is the component of $\smash{\hat{\mathbf{d}}}$ along the probe field, $\smash{\hat{\mathbf{d}}.\mathbfcal{E}_p = \hat{d}_p \mathcal{E}_p}$, and $\smash{\hat{P}_g}$ projects onto the ground manifold.
 
In the absence of Doppler shift, the zero-temperature susceptibility $\chi_0$ does not depend on the directions of the probe or the coupling lasers. At room temperature, however, one has to average $\hat{\rho}$ over a thermal distribution of atoms. An atom moving towards the probe at speed $v$ sees it blue shifted by $\omega_p v / c$. For the same atom the coupling frequency is red (blue) shifted by $\omega_c v / c$ if it is  counterpropagating (copropagating). Hence, for the two cases the thermal susceptibility is given by
\begin{equation}
    \chi = \frac{1}{\sqrt{\pi}} \int_{-\infty}^{\infty} \! {\rm d}u \; e^{-u^2} 
    \chi_0 \big( \Delta_p + \Delta_p^{T} u, 
    \Delta_c \mp \Delta_c^{T} u\big) \;,
    \label{eq:thermalaveraging}
\end{equation}
with $u = v/v_T$, $\smash{\Delta_{p(c)}^T \coloneqq \omega_{p(c)} v_{T} / c}$, and $v_{T} = \sqrt{2 k_B T / m}$, $m$ being the mass of a ${}^{87}$Rb atom and $k_B$ the Boltzmann constant. At room temperature $T=300$ K the Doppler shifts $\smash{\Delta_{p(c)}^T}$ are hundreds of MHz, which strongly rescale the susceptibility. However, since $v_T/c \sim 10^{-6}$ relativistic Doppler shifts are negligible.

To clearly observe the transmission peaks in the experiment, we subtract out a broad reference signal resulting from only the probe beam (see Fig.~\ref{fig:setup}). This corresponds to the difference susceptibility $\bar{\chi} \coloneqq \chi^{g,e} - \chi$, where $\chi^{g,e}$ is the susceptibility without the Rydberg levels. One can obtain $\bar{\chi}$ by thermally averaging its zero-temperature counterpart $\bar{\chi}_0 \coloneqq \chi^{g,e}_0 - \chi_0$ following Eq.~\eqref{eq:thermalaveraging}. Crucially, as a function of the probe detuning $\chi^{g,e}$ exhibits a single peak whose width is set by $\smash{\Delta_{p}^T} \sim 300$ MHz, whereas the linewidths in $\bar{\chi}$ are limited by the excited-state lifetime (tens of MHz). Thus, employing the EIT scheme allows us to resolve magnetic fields that are orders of magnitude smaller than what would be possible otherwise.

\section{Results and discussion}\label{sec:Measurements}
\begin{figure}
    \includegraphics[width=\columnwidth]{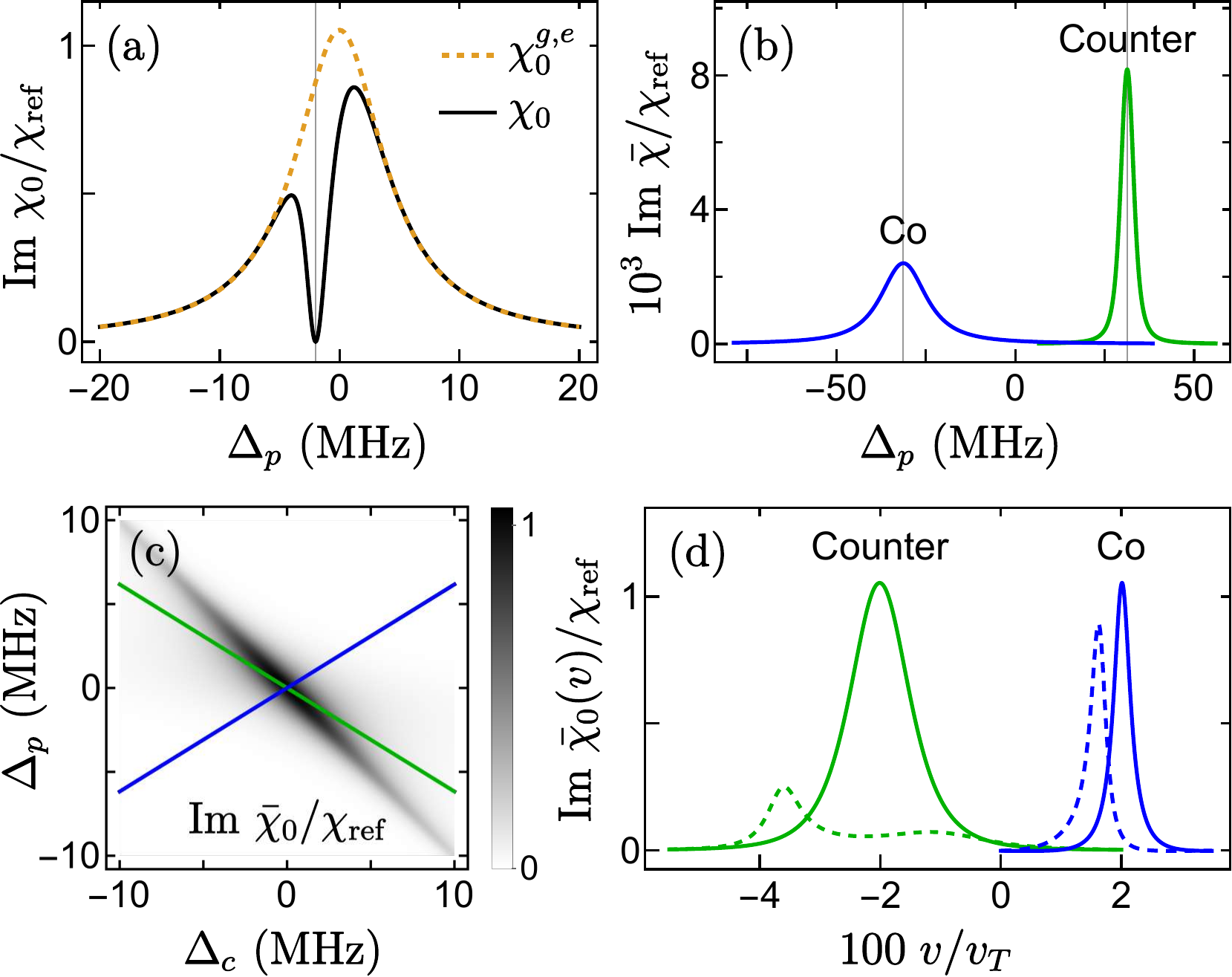}
    \centering
    \caption{
    (a) Imaginary part of the zero-temperature, steady-state susceptibility $\chi_0$ as a function of the probe detuning $\Delta_p$ for magnetic field $B=0$, coupling detuning $\Delta_c = 2$ MHz, and $\sigma^+$-$\sigma^-$ polarizations of the two lasers. The dotted curve corresponds to the susceptibility without the coupling beam. The vertical line shows two-photon resonance condition $\Delta_p = -\Delta_c$. Here $\chi_{\text{ref}} = 2 N e a_0 / (\varepsilon_0 \mathcal{E}_p) \approx 2 \times 10^{-4}$. 
    (b) Thermally averaged difference susceptibility, $\bar{\chi} \coloneqq \chi^{g,e} - \chi$, for counter- and copropagating lasers for the same setup with $\Delta_c = -51$ MHz. Vertical lines show the peak locations $\Delta_p = \mp \Delta^*$ where $\Delta^* \coloneqq (\omega_p / \omega_c) \Delta_c$ and $\omega_{p(c)}$ is the probe (coupling) frequency. (c) Zero-temperature difference susceptibility $\bar{\chi}_0$ showing a narrow EIT window. Green and blue lines represent thermal averaging for counter- and copropagating peaks, $\Delta_p = \mp \Delta^*$. 
    (d) Contribution of different velocities to the susceptibility, $\chi_0(v) \coloneqq \chi_0(\Delta_p + \omega_p v/c, \Delta_c \mp \omega_c v/c)$, for $\Delta_c = -10$ MHz at $\Delta_p = \mp \Delta^*$ (solid lines) and at $\Delta_p = \mp \Delta^*/2$ (dashed lines).
    }
    \label{fig:theory_circular}
\end{figure}

The simplest scenario that gives EIT is when the two lasers have opposite circular polarizations, e.g., probe $\sigma^+$ and coupling $\sigma^-$. Then, as sketched in Fig.~\ref{fig:energylevels}, the drive and loss bring the atoms to the manifold of $|m_F^g = 2 \rangle$, $|m_F^e = 3 \rangle$, and $|m_I^r = 3/2, m_J^r = 1/2 \rangle$, so the steady state reduces to that of a 3-level EIT \cite{EIT_Review_Marangos}. In particular, Im $\chi_0$ drops to zero at two-photon resonance, signaling a sharp transparency window [Fig.~\ref{fig:theory_circular}(a)]. Away from resonance $\chi_0$ matches the Lorentzian profile of Im $\smash{\chi_0^{g,e}}$ obtained without the coupling laser. Hence, the difference $\bar{\chi}_0$ has an imaginary part that is peaked at $\Delta_p = -\Delta_c$ and maximized for $\Delta_c = 0$ when $B=0$.

In thermally averaging $\bar{\chi}_0$ the largest weight comes from velocities $u^*$ for which the lasers are close to two-photon resonance, i.e., $|\Delta_p + \Delta_c + \big(\Delta_p^T \mp \Delta_c^T\big) u^*| < w_{\text{EIT}}$  [see Eq.~\eqref{eq:thermalaveraging}], where $w_{\text{EIT}}$ is the EIT window. For detunings of order tens of MHz, $u^* < 1$ and the integral can be approximated as
\begin{equation}
    \text{Im } \bar{\chi} \approx \frac{A}{\omega_c \mp \omega_p} \;
    \text{Im } \bar{\chi}_0^{g,e} \!\left( 
    \frac{\omega_c \Delta_p \pm \omega_p \Delta_c}{\omega_c \mp \omega_p}
    \right) ,
    \label{eq:difference_suscep}
\end{equation}
where $A$ is a slowly-varying function of $\Delta_p$ and $\Delta_c$. Here we assume $|\omega_c - \omega_p| \gg \omega_p w_{\text{EIT}} / w_{g,e}$, where $w_{g,e}$ is the two-level absorption window (width of $\smash{\chi_0^{g,e}}$), set by the lifetime of the excited state. For sharp EIT $w_{\text{EIT}} \ll w_{g,e}$, the assumption is usually well justified. From Eq.~\eqref{eq:difference_suscep} the transmission peak occurs at $\Delta_p = \mp (\omega_p/\omega_c) \Delta_c$ for the counter- and copropagating cases, respectively, with linewidths $(1 \mp \omega_p / \omega_c) w_{g,e}$ and heights $\propto 1/(\omega_c \mp \omega_p)$. So the peak is narrower and taller for counterpropagating lasers [see Fig.~\ref{fig:theory_circular}(b)]. This can be understood by noting that a broader range of velocities contribute to the signal in this configuration [Figs.~\ref{fig:theory_circular}(c, d)], but this range falls off rapidly as one deviates from $\Delta_p = - (\omega_p/\omega_c) \Delta_c$.

\begin{figure}
    \centering
    \includegraphics[width=\columnwidth]{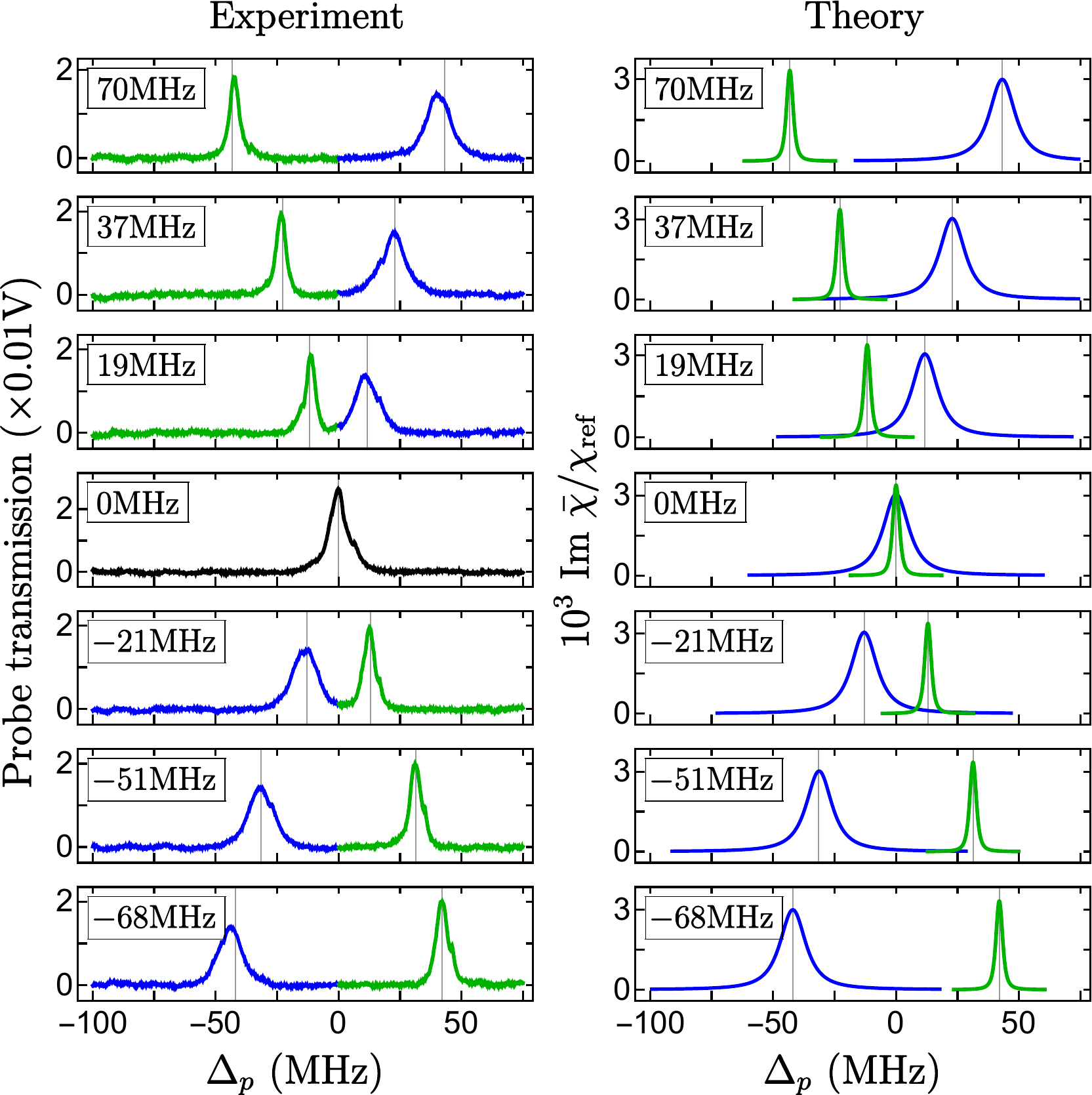}
    \caption{
    Measurements (left) and simulations (right) for different values of $\Delta_c$ and $B=0$ when the lasers are both linearly polarized along $x$ (perpendicular to $\mathbf{B}$). Green and blue curves show the counter- and copropagating cases, respectively. Vertical lines show the corresponding peak locations at $\Delta_p = \mp (\omega_p / \omega_c) \Delta_c$. For clarity the numerical susceptibility for the copropagating case is multiplied by a factor of 3 (see text). Note, all parameter values are listed in Appendix~\ref{app:parameters}.
    }
    \label{fig:detuning}
\end{figure}

Figure~\ref{fig:detuning} shows a side-by-side comparison of experimental measurements and numerical simulations with $B=0$ for the case where both lasers are linearly polarized. Here the physics does not reduce to three levels as either laser can induce $\sigma^+$ or $\sigma^-$ transitions (see Fig.~\ref{fig:energylevels}). However, as the sublevels in each manifold are degenerate, the transmission spectrum is similar: In both theory and experiment we observe a narrow peak at $\Delta_p = - (\omega_p/\omega_c) \Delta_c$ for counterpropagating lasers and a relatively broad peak at $\Delta_p = (\omega_p/\omega_c) \Delta_c$ for copropagating lasers. This is seen for a wide range of blue and red detuning.

The main difference between the measurements and simulations is that the relative signal height for the copropagating case is smaller in the simulations. (Note that in the experiment the power of the coupling laser was adjusted to match the heights for the two configurations.) This discrepancy originates from the fact that as the coupling power is reduced the experimental signals go to zero but the steady-state susceptibilities do not (see Fig.~\ref{fig:tx_vs_power} in Appendix~\ref{app:signalvspower}). This is because the relaxation time diverges and effects such as dephasing due to laser linewidths and stochastic resetting due to atoms leaving the probe beam come into play \cite{2017_Spreeuw}. Note, however, these do not affect the locations or widths of the peaks.

\begin{figure}
    \centering
    \includegraphics[width=\columnwidth]{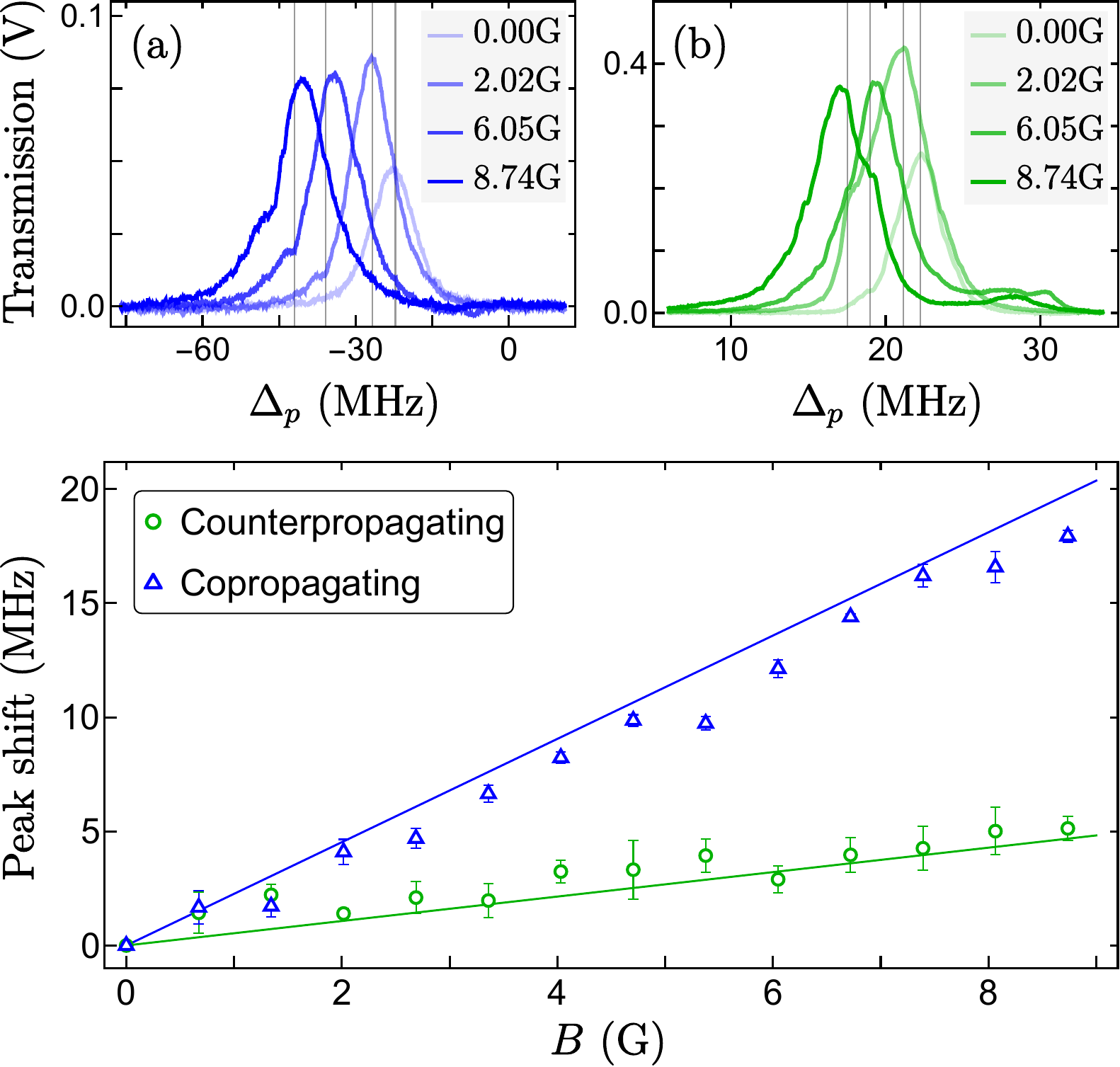}
    \caption{
    Top: Measurements showing the variation of the EIT signal with $B$ for $\sigma^-$-$\sigma^+$ polarizations and $\Delta_c = -36$ MHz for (a) copropagating and (b) counterpropagating lasers. Vertical lines show the predicted peak locations at $\Delta_p(-B)$ given by Eq.~\eqref{eq:peakloc_circular}. Bottom: Displacement of the fitted peaks with $B$. Solid lines show theoretical predictions.
    }
    \label{fig:circular}
\end{figure}

\begin{figure}[b]
    \includegraphics[width=\columnwidth]{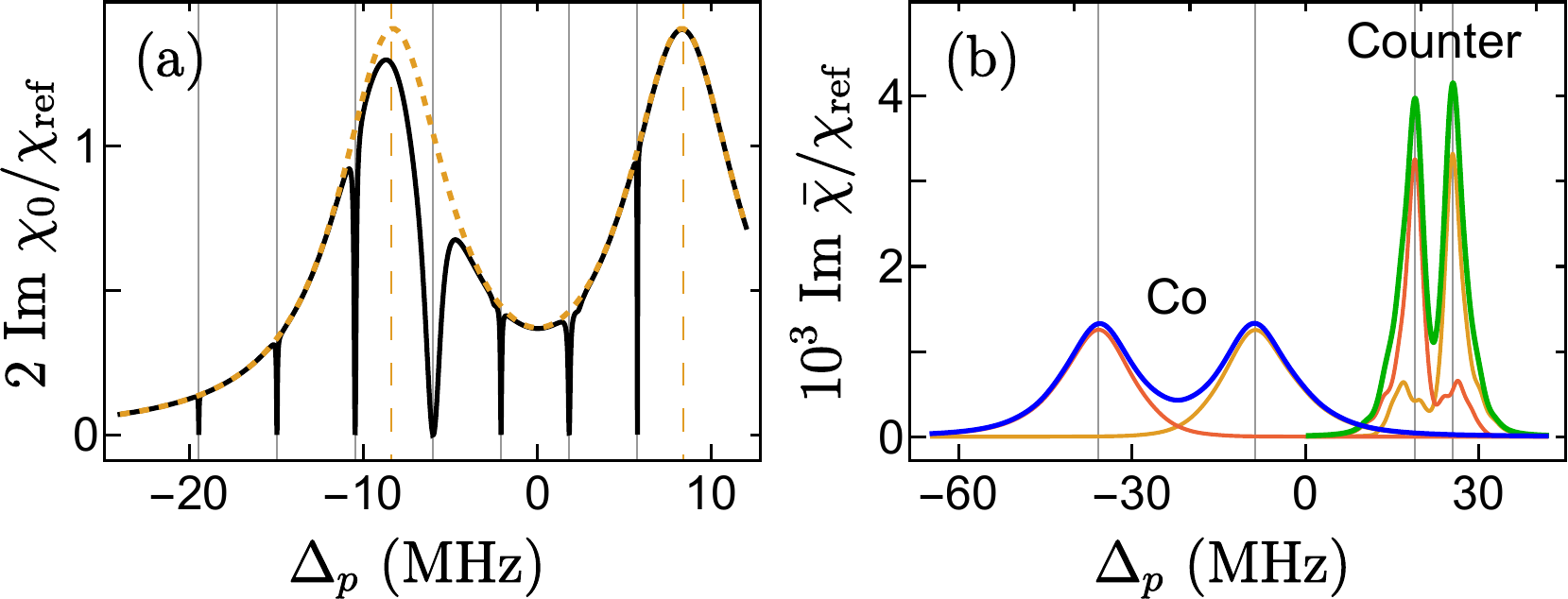}
    \centering
    \caption{
    (a) Zero-temperature susceptibility (solid curve) for $x$-$x$ polarizations with $B = 6$ G and $\Delta_c = 6$ MHz, showing multiple EIT windows at two-photon resonances between a ground level and a Rydberg level with the same $m_F$ (solid vertical lines). The dotted curve shows $\chi_0^{g,e}$, exhibiting peaks at $\Delta_p = \pm \mu_B B$ (dashed vertical lines). (b) Thermally averaged difference susceptibility for the same setup with $\Delta_c = -36$ MHz, exhibiting two peaks corresponding to the $\sigma^{\pm}$-$\sigma^{\mp}$ and  polarizations, $\Delta_p(\pm B)$ in Eq.~\eqref{eq:peakloc_circular} (vertical lines). Yellow and red curves show the main contributions to $\bar{\chi}$ from coherences between the $|m^g_F = \pm 2\rangle$ and $|m^e_F = \pm 3\rangle$ levels, respectively.
    }
    \label{fig:theory_xx}
\end{figure}

The response to a magnetic field $B$ is again easiest to understand for the $\sigma^+$-$\sigma^-$ configuration. Here, $B$ shifts the ground level $|m_F^g = 2 \rangle$ by $\Delta_g = g^g_F m^g_F \mu_B B = \mu_B B$, the excited level $|m_F^e = 3 \rangle$ by $\Delta_e = g^e_F m^e_F \mu_B B = 2 \mu_B B$, and the Rydberg level $|m_I^r = 3/2, m_J^r = 1/2 \rangle$ by $\Delta_r \approx g^r_J m^r_J \mu_B B = \mu_B B$ (Fig.~\ref{fig:energylevels}). Hence, the effective detuning of the probe is $\smash{\Delta_p^{\text{eff}}} = \Delta_p + \Delta_g - \Delta_e = \Delta_p - \mu_B B$ and that of the coupling is $\smash{\Delta_c^{\text{eff}}} = \Delta_c + \Delta_e - \Delta_r \approx \Delta_c + \mu_B B$. The condition for a transmission peak is the same as before, namely, $\smash{\Delta_p^{\text{eff}}} = \mp (\omega_p/\omega_c) \smash{\Delta_c^{\text{eff}}}$, which gives
\begin{equation}
    \Delta_p (B) \approx \mp (\omega_p/\omega_c) \Delta_c + (1 \mp \omega_p/\omega_c) \mu_B B .
    \label{eq:peakloc_circular}
\end{equation}
Therefore, as $B$ is varied the peak for the copropagating case moves faster by a factor $(\omega_c + \omega_p) / (\omega_c - \omega_p) \approx 4.2$. The same is true for a $\sigma^-$-$\sigma^+$ configuration with $B$ replaced by $-B$. This enhanced response is shown in Fig.~\ref{fig:circular}. Note that, in general, the peak varies linearly with a slope
\begin{equation}
    \frac{\partial\Delta_p}{\partial(\mu_B B)} \approx 
    g^e_F m^e_F - g^g_F m^g_F \mp \frac{\omega_p}{\omega_c} (g^e_F m^e_F - g^r_J m^r_J) \;.
    \label{eq:peakvariation}
\end{equation}
Thus, copropagating lasers give an advantage whenever the Zeeman shift of the excited level is higher (or lower) than those of both the ground and the Rydberg levels. 

\begin{figure}[b]
    \centering
    \includegraphics[width=\columnwidth]{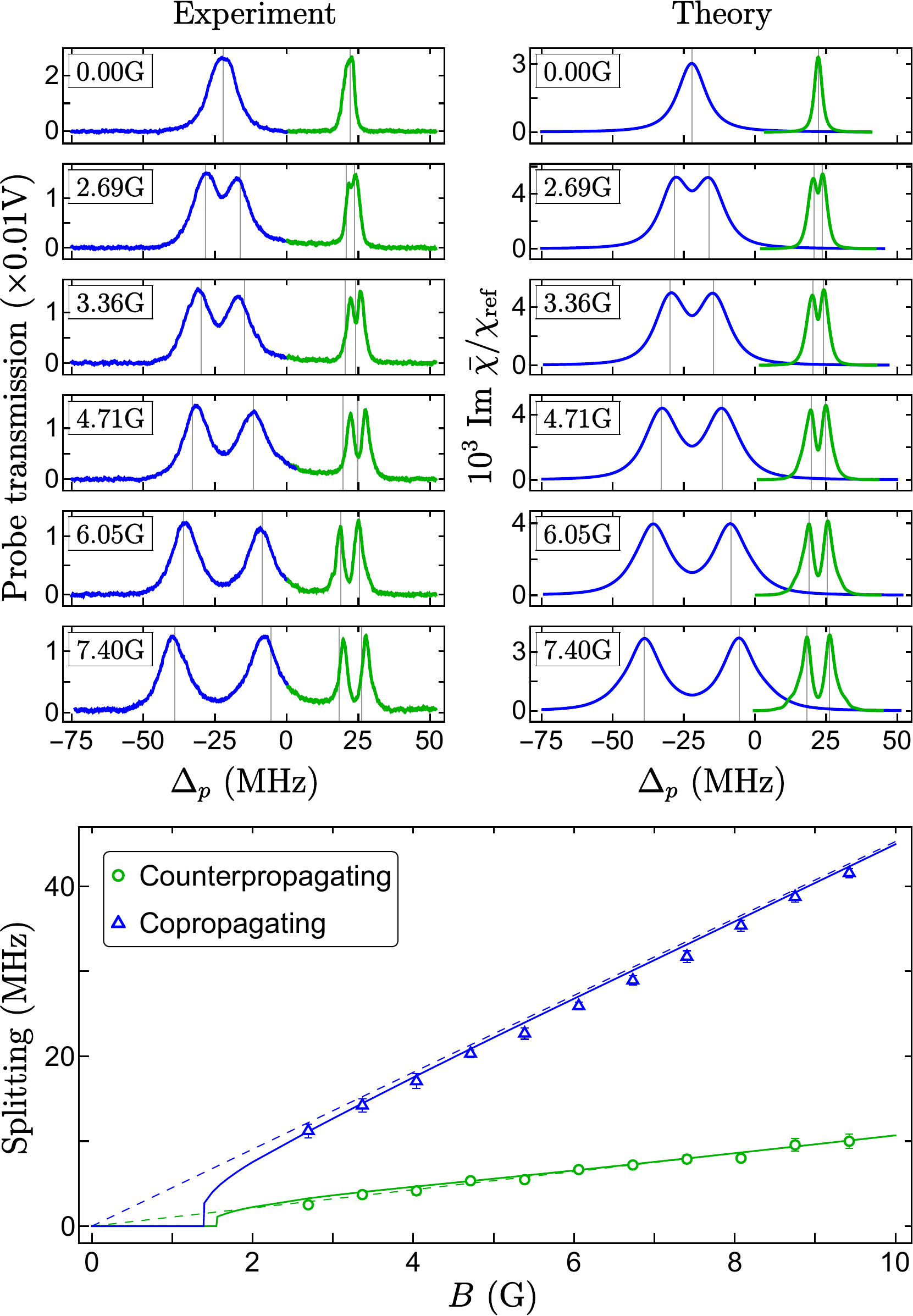}
    \caption{
    Top: Measurements (left) and simulations (right) for increasing $B$ with $\Delta_c = -36$ MHz and $x$-$x$ polarizations. As before, green and blue curves stand for counter- and copropagating cases. Vertical lines show $\Delta_p(\pm B)$ from Eq.~\eqref{eq:peakloc_circular}. Measurements are averaged over $x$-$x$ and $y$-$y$ polarizations to correct for any misalignment of lasers. As in Fig.~\ref{fig:detuning}, $\bar{\chi}$ for the copropagating case is enlarged by a factor of 3. Bottom: Fitted peak separation as a function of $B$, showing an enhanced response for copropagating lasers. Solid lines are numerical fits and dashed lines are predicted using Eq.~\eqref{eq:peakloc_circular}.
    }
    \label{fig:linear}
\end{figure}

When the lasers are linearly polarized, different paths connecting the ground and Rydberg manifolds acquire different Zeeman shifts (see Fig.~\ref{fig:energylevels}), which splits the corresponding two-photon resonances, producing multiple transmission peaks at zero temperature. This is shown in Fig.~\ref{fig:theory_xx}(a), where we find numerically that Im $\chi_0$ drops to zero whenever the resonant levels satisfy $m^r_I + m^r_J = m^g_F$ and the lasers have identical polarization. Upon thermal averaging, the difference susceptibility $\bar{\chi}$ generically has a two-peaked Doppler-broadened profile as in Fig.~\ref{fig:theory_xx}(b). For polarizations perpendicular to the magnetic field, as in our experimental setup, the response is dominated by extreme $m_F$ values corresponding to the $\sigma^+$-$\sigma^-$ and $\sigma^-$-$\sigma^+$ configurations. Hence, from Eq.~\eqref{eq:peakloc_circular} the peak separation is given by $\Delta_{pp} \approx 2 (1 \mp \omega_p/\omega_c) \mu_B B$ for counter- and copropagating cases, respectively.

Figure~\ref{fig:linear} shows experimental measurements which corroborate these predictions. We observe an order of magnitude bigger splitting in the transmission spectrum for copropagating lasers. The peak separations, plotted in the lower panel, match numerical simulations and are indeed magnified by the factor $(\omega_c + \omega_p) / (\omega_c - \omega_p) \approx 4.2$ relative to the counterpropagating setup. We attribute the slight mismatch between the predicted and measured peak locations for the counterpropagating case to a slow frquency drift of the lasers. In Appendix~\ref{app:otherpol} we present simulations that show for other polarizations the spectrum can also exhibit multiple peaks.

\section{Conclusion and Outlook} \label{sec:Conclusion_Outlook}
We have performed quantum magnetometry with thermal ${}^{87}$Rb atoms using a Rydberg EIT scheme. We have demonstrated that, contrary to the prevailing custom of using counterpropagating lasers to mitigate Doppler effects, one can harness such effects in a copropagating arrangement to obtain a greatly enhanced response to magnetic fields. The enhancement depends on the ratio of the laser frequencies, which can be varied by targeting different energy levels. Our measurements are in good agreement with theoretical modeling based on a Markovian master equation, and pave the way to precision magnetometry in room-temperature setups. 

For simplicity we have focused exclusively on a one-dimensional geometry. It would be interesting to see if varying the relative orientations of the magnetic field and the probe and coupling beams produces more dramatic effects. A limitation of the current approach is that the transmission peaks are broader for the copropagating case, which limits the resolution. Thus, it would be very useful to explore if there are ways to control the linewidth independently of the peak position. Our work provides strong motivations for these developments which can lead to important applications in quantum sensing with Rydberg atoms and \cite{2019_Haroche} and possibly magnetometry beyond the standard quantum limit \cite{2015_PRX}.

\section*{Acknowledgments}
S.R. acknowledges funding from the Department of Science and Technology, India, via the WOS-A project grant no. SR/WOS-A/PM-59/2019. We acknowledge the contribution of Meena M. S. and the RRI mechanical workshop for their assistance with the experiments. S.D. acknowledges the hospitality of the Max Planck Institute for the Physics of Complex Systems, where part of the calculations was done. S.K.B. acknowledges the funding from the I-HUB Quantum Technology Foundation via the SPIKE Project grant no. I-HUB/SPIKE/2023-24/004.

\appendix
\section{Simulations for other polarizations of the probe and the coupling lasers}
\label{app:otherpol}

\begin{figure}[h]
    \centering
    \includegraphics[width=\columnwidth]{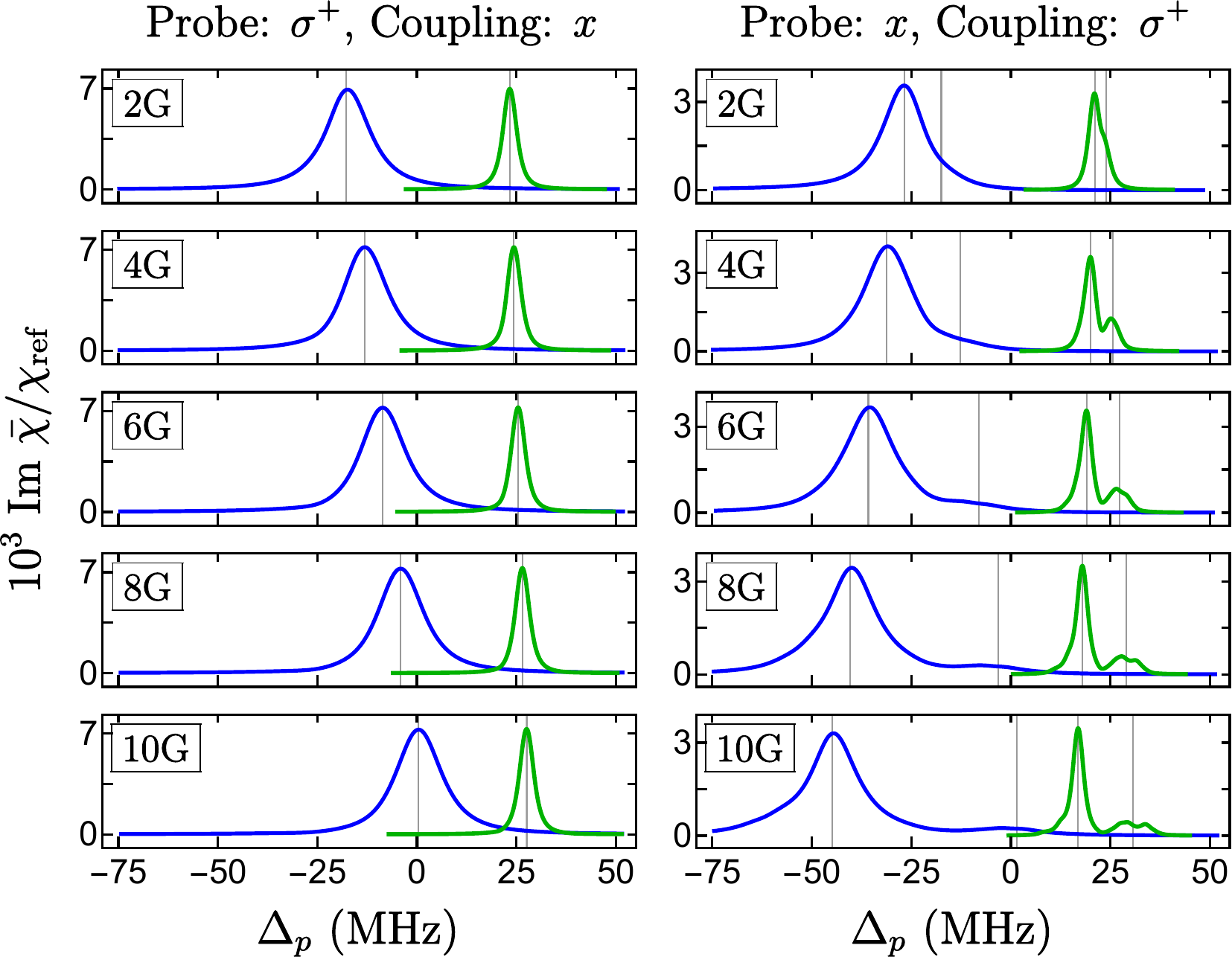}
    \caption{
    Steady-state difference susceptibility with $\Delta_c = -36$ MHz for $\sigma^+$-$x$ (left) and $x$-$\sigma^+$ (right) polarizations. For $\sigma^+$-$x$ the peak follows $\Delta_p(B)$ from Eq.~\eqref{eq:peakloc_circular}. For $x$-$\sigma^+$ the dominant peak follows $\Delta_p(-B)$; the other line is obtained from Eq.~\eqref{eq:peakvariation} with $(m^g_F, m^e_F, m^r_J) = (-1, -2, 1/2)$ for counterpropagating (green) and $(0, 1, -1/2)$ for copropagating (blue) geometries.
    }
    \label{fig:1x_x1}
\end{figure}

In Fig.~\ref{fig:1x_x1} we show the response for the $\sigma^+$-$x$ and $x$-$\sigma^+$ polarizations. For the former the steady state is almost identical to that of the $\sigma^+$-$\sigma^-$ case discussed in the main text. Hence, both counter- and copropagating configurations exhibit a single peak given by Eq.~\eqref{eq:peakloc_circular}. For the latter the steady state is close to that of the $\sigma^-$-$\sigma^+$ case, so we see a single dominant peak in both configurations; however, we can also discern a satellite peak for the counterpropagating case resulting from other two-photon resonances. Note that such a peak was observed in the experiment for the $\sigma^-$-$\sigma^+$ setting [see Fig.~\ref{fig:circular}(b)], which we attribute to an imperfect alignment of the lasers.

\begin{figure}[h]
    \centering
    \includegraphics[width=\columnwidth]{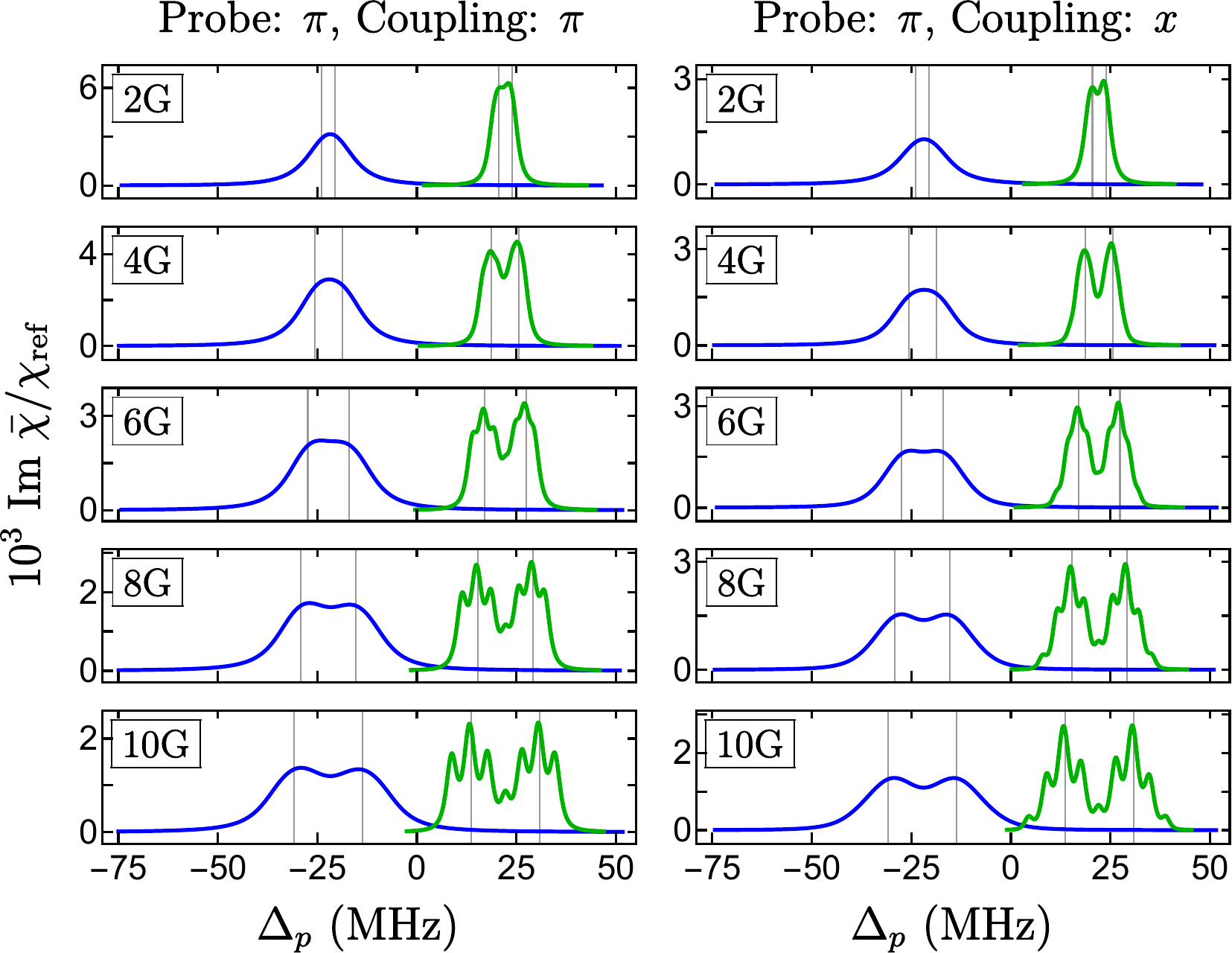}
    \caption{
    Steady-state difference susceptibility with $\Delta_c = -36$ MHz for $\pi$-$\pi$ (left) and $\pi$-$x$ (right) polarizations. For both counterpropagating (green) and copropagating (blue) cases the vertical lines tracking the peaks are obtained from Eq.~\eqref{eq:peakvariation} with $(m^g_F, m^e_F, m^r_J) = (0, 0,\pm 1/2)$.
    }
    \label{fig:zz_zx}
\end{figure}

Figure~\ref{fig:zz_zx} shows when the probe is $\pi$ polarized (parallel to the magnetic field) and the coupling laser is linearly polarized, the counterpropagating arrangement produces multiple peaks whereas the copropagating case exhibits only two shallow peaks. Now the dominant contribution comes from paths with $m^g_F = 0$, $m^e_F = 0$ and $m^r_J = \pm 1/2$ (cf.~Fig.~\ref{fig:energylevels}), hence from Eq.~\eqref{eq:peakvariation} the central peak separation is comparable in the two cases. This example illustrates the response can be sensitive to the polarizations.

\section{Signal strength vs coupling laser power}
\label{app:signalvspower}

\begin{figure}[h]
    \centering
    \includegraphics[width=\columnwidth]{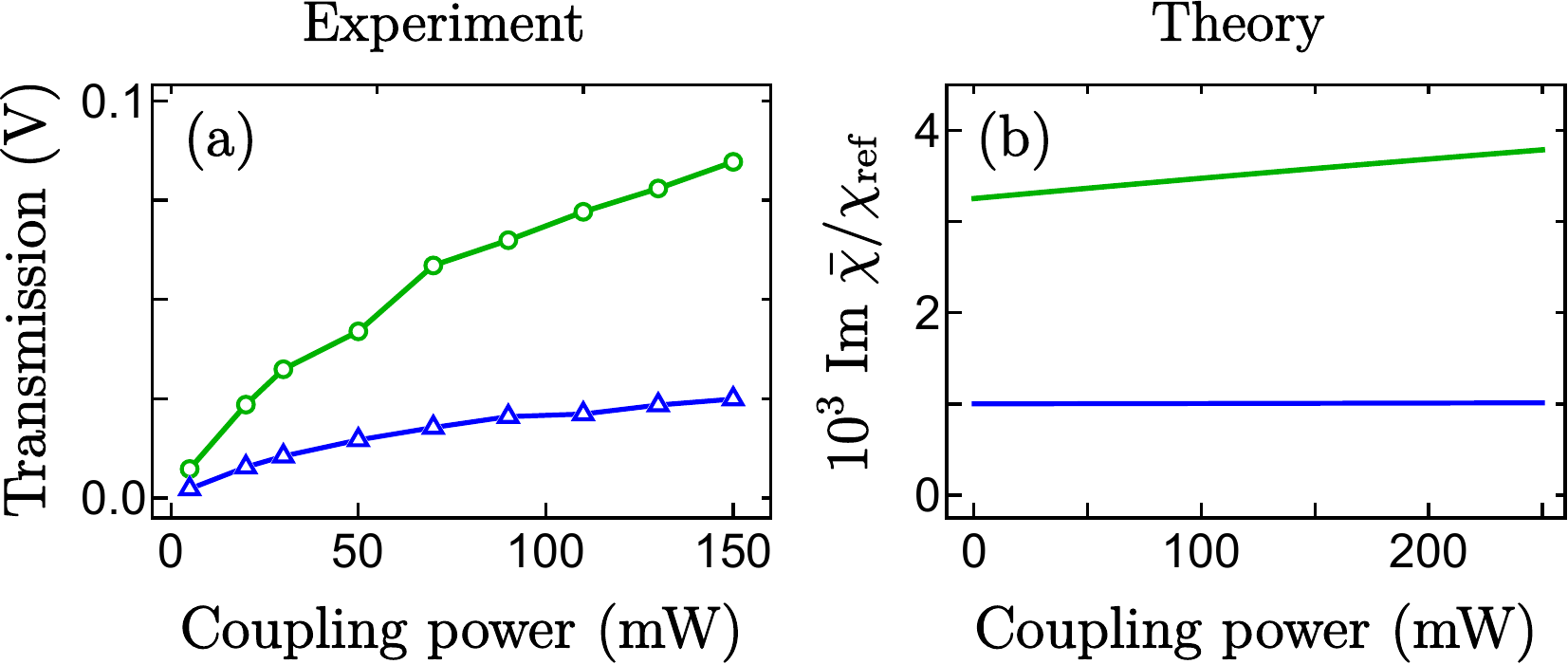}
    \caption{
    Measurements (left) and simulations (right) of the peak EIT signal strength as a function of the coupling laser power for $B=0$, $\Delta_c = -36$ MHz, and $x$-$x$ polarizations.
    }
    \label{fig:tx_vs_power}
\end{figure}

Figure~\ref{fig:tx_vs_power} shows how the EIT signal strength changes as a function of the coupling laser power ($P_c$), which enters the Rabi frequencies between the excited and the Rydberg levels. While the experimental signal vanishes for $P_c \to 0$ in both counter- and copropagating configurations, the numerically computed steady-state susceptibilities remain largely unaltered. As discussed in the main text, this mismatch arises from a diverging relaxation time at weak coupling, where slow processes such as dephasing and stochastic resetting become significant.

\section{Parameter values used for measurements and numerical simulations}
\label{app:parameters}

The power and beam radius of the probe laser were set to 0.015 mW and 699 $\mu$m, respectively. For the coupling laser we used a beam radius of 731.5 $\mu$m for copropagating arrangements and 1227 $\mu$m for counterpropagating arrangements as well as for the zero-temperature calculations in Figs.~\ref{fig:theory_circular} and \ref{fig:theory_xx}. Table~\ref{tab:Pc} lists the values of the coupling laser power used for the different cases.

\begin{table}[h]
    \renewcommand{\arraystretch}{1.2}
    \caption{\label{tab:Pc}Coupling laser power $P_c$ used in different plots.}
    \begin{ruledtabular}
    \begin{tabular}{l|l|r}
    Figure \hspace*{0.6cm} & Arrangement & $P_c$ (mW) \\
    \hline 
    Figure~\ref{fig:theory_circular} & $T=0$ & 220\\
    Figure~\ref{fig:detuning} & Counterpropagating \hspace*{0.5cm} & 50\\
    Figure~\ref{fig:detuning} & Copropagating & 250\\
    Figure~\ref{fig:circular} & Counterpropagating & 220\\
    Figure~\ref{fig:circular} & Copropagating & 220\\
    Figure~\ref{fig:theory_xx} & $T=0$ & 20\\
    Figure~\ref{fig:linear} & Counterpropagating & 20\\
    Figure~\ref{fig:linear} & Copropagating & 150\\
    Figure~\ref{fig:1x_x1} & Counterpropagating & 20\\
    Figure~\ref{fig:1x_x1} & Copropagating & 150\\
    Figure~\ref{fig:zz_zx} & Counterpropagating & 20\\
    Figure~\ref{fig:zz_zx} & Copropagating & 150\\
    Figure~\ref{fig:tx_vs_power} & Counterpropagating & 20\\
    Figure~\ref{fig:tx_vs_power} & Copropagating & 150\\
\end{tabular}
\end{ruledtabular}
\end{table}

\bibliography{ref}

\end{document}